FRONT MATTER

## Title

**Cooperative rheological state-switching of enzymatically-driven composites of circular DNA and dextran**

Short title: State-switching of topologically-active DNA composites

## Authors

Juexin Marfai, Ryan J. McGorty, Rae M. Robertson-Anderson*

## Affiliations

Department of Physics and Biophysics, University of San Diego, San Diego, CA 92110
*randerson@sandiego.edu

## Abstract

Polymer topology, which plays a principal role in the rheology of polymeric fluids, and non-equilibrium materials, which exhibit time-varying rheological properties, are topics of intense investigation. Here, we push composites of circular DNA and dextran out-of-equilibrium via enzymatic digestion of DNA rings to linear fragments. Our time-resolved rheology measurements reveal discrete state-switching, with composites undergoing abrupt transitions between dissipative and elastic-like states. The gating time and lifetime of the elastic-like states, and the magnitude and sharpness of the transitions, are surprisingly decorrelated from digestion rates and non-monotonically depend on the DNA fraction. We model our results using sigmoidal two-state functions to show that bulk state-switching can arise from continuous molecular-level activity due to the necessity for cooperative percolation of entanglements to support macroscopic stresses. Our platform, coupling the tunability of topological composites with the power of enzymatic reactions, may be leveraged for diverse material applications from wound-healing to self-repairing infrastructure.

## Teaser

Non-equilibrium DNA-dextran composites exhibit bulk rheological switching via continuous enzymatic fragmentation of ring DNA.

# MAIN TEXT

## Introduction

The interiors of biological cells are out-of-equilibrium, crowded composite materials comprising high concentrations of interacting macromolecules, such as DNA and proteins, that are continuously restructuring via the action of enzymes and molecular motors. These features of biological systems–crowded, composite, and non-equilibrium–give rise to unique viscoelastic properties that are time-dependent and tunable. For example, myriad naturally occurring restriction enzymes convert densely packed circular DNA chains to linear topology, and cleave long linear chains into shorter fragments (*1*, *2*). These topological conversions, critical to diverse processes such as transcription and repair (*3*), change the nature of the intra- and inter-chain interactions (e.g., entanglements) and can drive changes to the viscoelastic properties of the crowded cellular environment (*1*, *4*, *5*). However, how microscale properties translate to bulk macroscopic rheological properties remains an open question that is critical to leveraging biological multifunctionality and non-equilibrium processes for materials applications.

For example, circular (ring) polymers, with no free ends, are unable to form the same types of entanglements with neighboring chains as their linear counterparts, limiting their ability to exhibit the elastic-like or rubbery rheology predicted by the reptation tube model that describes entangled linear polymers (*6–9*). Because there is no straightforward way to extend the reptation model to ring polymers, as it relies on the free ends of linear chains, theories such as the fractal loopy globule (FLG) model have been proposed that describe the conformation of an 'entangled' ring as a self-similar fractal of loops (*10*). FLG redefines the concepts of an entanglement tube radius and polymer length between entanglements in terms of the average spacing between topological constraints (*10*).

Moreover, threading of rings by other rings and linear chains has been shown to increase the viscosity, elastic plateau modulus, and relaxation timescales of ring-linear blends, as compared to their pure ring and linear counterparts (*8*, *11–17*). Simulations of entangled ring polymers have also demonstrated that above a critical polymer concentration and length, ring-ring threadings can lead to glassy behavior that is characterized by heterogeneous chain dynamics with discrete clusters of 'slow' threaded rings and 'fast' unthreaded or minimally threaded rings (*15*). These studies further showed evidence of cooperativity between clustered rings that was generically analogous to other glassy systems (polymers, complex fluids, colloids, etc.) in which the glass transition occurs when the average size of the 'slow' clusters reaches the percolation threshold, i.e., they span the system size (*18*, *19*). A universal indicator of the heterogeneous dynamics and clustering of components with different mobilities expected for systems near a glass transition has been shown to be non-Gaussian distributions of the displacements of the constituents (e.g., polymers, colloids, etc), also termed van Hove distributions (*20*). While Gaussian van Hove distributions are expected for spatiotemporally homogeneous dynamics, for systems which have a combination of slow- and fast-moving populations, e.g., near a glass transition, van Hove distributions are expected to exhibit exponential large-displacement tails that extend beyond the Gaussian profile due to the broad spectrum of mobilities of the fast-moving particles. Glassy distributions may also exhibit higher probabilities of near-zero displacements owing to the slow, kinetically 'caged' components.

Recent microrheology measurements have demonstrated that concentrated circular DNA solutions steadily become increasingly viscous under the continuous action of enzymes that cut the polymers once, converting them to linear topology (*21*). The viscous thickening of such 'topologically-active' DNA solutions during linearization was shown to arise from the larger radius of gyration $R_G$ of linear chains compared to that of circular polymers, which increases the polymer coil overlap. Conversely, the same study showed that solutions of entangled linear DNA gradually becomes less viscous over time under the action of restriction enzymes that cut the long chains into many short fragments (*21*). This effect was rationalized as arising from a decreasing number of entanglements per chain as the lengths of the chain fragments $l_f$ become comparable to or lower than the nominal length between entanglements $l_e$. In these studies, the enzymes act as catalysts to break the sugar-phosphate DNA backbone at specific sites, thereby irreversibly altering the DNA, which, in turn, pushes the systems out-of-equilibrium, driving them to new thermodynamic equilibria. In other words, during enzymatic digestion, the systems are en route between two distinct equilibrium states, resulting in time-dependent variations in their rheological properties.

The propensity for entanglements and threading have also been shown to play an important role in composites of DNA and other biological and synthetic polymers, including microtubules (*22–24*) and dextran (*25, 26*). Previous experiments examining the rheological properties of viscoelastic composites of DNA and dextran polymers reported surprising non-monotonic dependences of rheological properties on the fraction of DNA in the composites $\phi_{DNA}$. For example, composites exhibited a higher elastic plateau modulus $G^0$ and reduced dissipation compared to either pure dextran or pure DNA solutions (*25*). This emergent topology-dependent behavior was shown to arise from DNA rings being either compacted by dextran (for $\phi_{DNA} \leq 0.5$) or swollen and threaded (for $\phi_{DNA} = 0.75$). Conversely, depletion-driven entropic stretching and increased self-association of linear DNA served to reduce connectivity at low $\phi_{DNA}$ or result in more densely entangled bundles at high $\phi_{DNA}$.

The results described above highlight the surprising rheological signatures that crowded polymeric systems, composites of different types of polymers, and topologically-active DNA fluids exhibit. Here, we seek to understand the interplay of these bio-inspired conditions by measuring the bulk rheological properties of composites of ring DNA and dextran in the presence of enzymes that linearize and fragment the DNA rings (Fig 1).

More generally, we aim to demonstrate that we can use a biological material and process (e.g., topologically-active DNA) to drive changes in a synthetic material (e.g., dextran). The use of dextran over another synthetic constituent is motivated by the known rheology and transport properties of steady-state DNA-dextran blends, facilitating reliable interpretation of non-equilibrium results. Moreover, dextran is the size of typical small soluble proteins in the cell which crowd much larger biopolymers such as DNA, and is often used as a model crowder in *in vitro* experiments aimed at understanding intracellular dynamics and processes (*26–30*). As such, this system may provide insights into how cellular crowding impacts enzymatic processes such as DNA digestion.

**Results**

We fix the overall polymer concentration to $11c^*$ and vary the volume fractions of $11c^*$ solutions of DNA ($\phi_{DNA}$) and dextran ($\phi_{dx} = 1 - \phi_{DNA}$) (Fig 1A). We measure the frequency-dependent viscoelastic moduli $G'(\omega)$ and $G''(\omega)$ of topologically-active composites with and without enzymatic activity (Fig 1D), and characterize the time-course of viscoelastic properties of the topologically-active composites en route from undigested to digested states (Fig 1E).

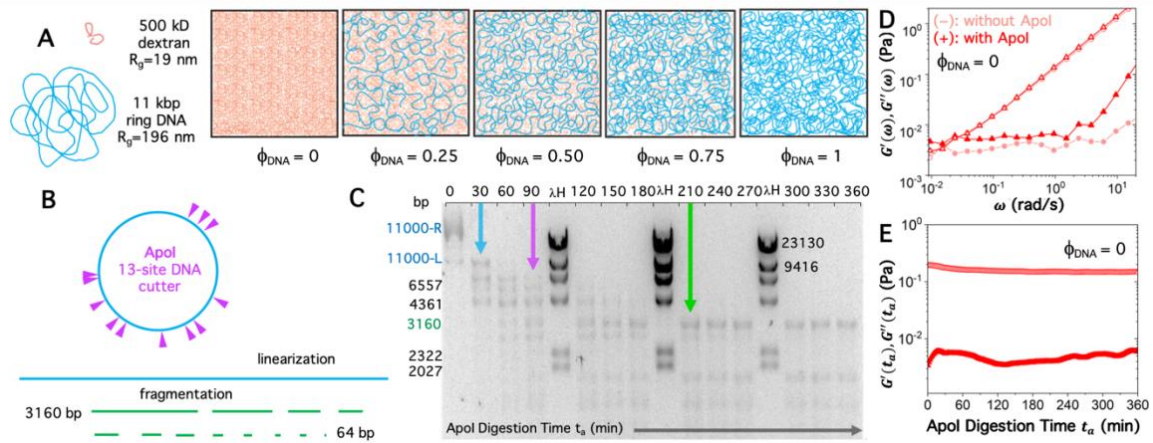

**Figure 1. Measuring the time-varying rheological properties of composites of dextran and topologically-active circular DNA.** (**A**) 11-kbp relaxed circular DNA ($R_G \simeq 196$ nm, blue) and 500 kDa dextran ($R_G \simeq 19$ nm, red) comprise $11c^*$ composites denoted by the DNA volume fraction: $\phi_{DNA} = 0$, 0.25, 0.5, 0.75, and 1. (**B**) We incorporate restriction endonuclease ApoI (magenta) into composites to cut DNA rings into 13 linear fragments with lengths of 64 to 3160 bp (green). (**C**) Time-resolved agarose gel electrophoresis shows the digestion of the $\phi_{DNA} = 1$ solution as described in Methods. Each lane shows the topological state of the DNA at a given time $t_a$ following the addition of ApoI. Lambda-HindIII markers ($\lambda H$) indicate the length of linear DNA that corresponds to each band, listed in basepairs in black font. The blue labels indicate 11-kbp ring (11000-R) and linear (11000-L) topologies, and the green label (3160) indicates the largest fragment following complete digestion. The blue arrow indicates complete linearization (11000-R band is no longer visible); the purple arrow indicates initial fragmentation of the full-length linear DNA (11000-L band is no longer visible); and the green arrow indicates complete digestion (the longest DNA length visible is 3160). We use a DHR3 rheometer to measure the (**D**) frequency-dependent elastic modulus $G'(\omega)$ (filled symbols) and viscous modulus $G''(\omega)$ (open symbols) of composites after 6 hrs with (+, dark shades) or without (-, light shades) ApoI, over three decades ($\omega = 0.009 - 20$ rad/s); and (**E**) time-dependent moduli $G'(t_a)$ and $G''(t_a)$ at $\omega = 1$ rad/s over 6 hours of ApoI digestion. Example data shown in (D) and (E) is for $\phi_{DNA} = 0$ which exhibits Newtonian fluid properties and no $t_a$ dependence.

***In situ fragmentation of entangled circular DNA induces an abrupt increase in bulk viscoelasticity.*** We first examine the effect of linearization and fragmentation of circular 11 kbp DNA on the bulk rheology of concentrated ($11c^*$) DNA solutions (Fig 1). This change in the topology and length of the DNA molecules is due to digestion by restriction endonuclease ApoI which cuts each circular plasmid into 13 linear fragments of different lengths ranging from 64 basepairs (bp) to 3.16 kilobasepairs (kbp) over the course of ~270 mins (Fig 1B,C). As seen in Fig 1C, and described in Methods, we purposefully optimized the ApoI:DNA stoichiometry to ensure that the digestion time $t_d$ is long compared to the measurement time $t_m$, such that we can treat the systems as in quasi-steady state on the timescale of each measurement. Specifically, at our chosen ApoI:DNA stoichiometry of 0.05 U/μg, the digestion time $t_d \gtrsim 270$ min is nearly two orders of magnitude longer than the measurement time, i.e., $\chi \simeq t_d/t_m \sim 10^2$.

To verify that the changes in bulk viscoelasticity of the DNA solutions (Fig 2A) are due to specific interactions between the DNA and ApoI, we perform control measurements for $11c^*$ dextran solutions (no DNA, $\phi_{DNA} = 0$) with and without ApoI. As expected, we find that dextran solutions exhibit viscous Newtonian dynamics (e.g., $G'' > G'$, $G''(\omega) \sim \omega$) that are similar with and without ApoI (Fig 1D) and independent of $t_d$ (Fig 1E).

Figure 2 compares the frequency-dependent elastic and viscous moduli $G'(\omega)$ and $G''(\omega)$ of the $11c^*$ DNA solution after 6 hours of ApoI digestion to an identical solution but without ApoI. The frequency dependence of $G'(\omega)$ and $G''(\omega)$ is similar with and without digestion, with both solutions exhibiting entanglement dynamics in which $G'(\omega)$ displays minimal $\omega$-dependence and $G'(\omega) > G''(\omega)$ across the entire frequency range (Fig 2A). However, ApoI digestion generally increases $G'$ and $G''$ (Fig 2A) and decreases the loss tangent $\tan\delta(\omega) = G''/G'$ (SI Fig S1).

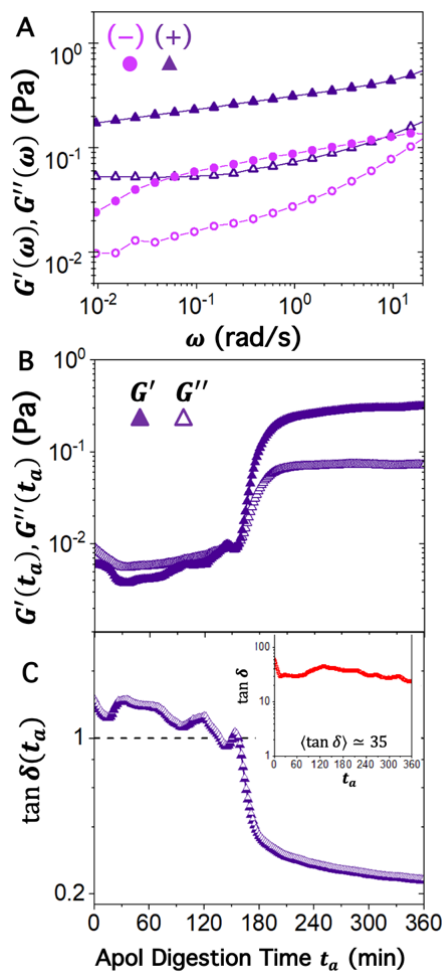

**Figure 2. Enzymatic digestion induces an abrupt increase in viscoelastic moduli of entangled ring DNA.** (**A**) $G'(\omega)$ (filled symbols) and $G''(\omega)$ (open symbols) of the $\phi_{DNA} = 1$ solution measured after 6 hours of either ApoI digestion (purple) or under identical conditions but with no ApoI (magenta) shows that ApoI-driven topological conversion increases the viscoelasticity of the circular DNA solution by an order of magnitude. (**B**) Time-dependent viscoelastic moduli $G'(t_a)$ (filled symbols) and $G''(t_a)$ (open symbols) for $\phi_{DNA} = 1$ during ApoI activity shows two distinct viscoelastic states, with $G'(t_a)$ sharply increasing by an order of magnitude at $t_a \simeq 150$ min. (**C**) Time-dependent loss tangent $\tan\delta(t_a)$ computed from the moduli shown in (**B**) with the horizontal dashed line indicating $G' = G''$. The inset shows that $\tan\delta(t_a)$ for the pure dextran solution ($\phi_{DNA} = 0$) is time-independent and substantially higher (more dissipative) than for pure DNA.

To understand these trends, we recall that $G'$ and $G''$ are measures of the elastic (energy storage) and viscous (energy dissipation) contributions to the stress response, and $\tan\delta$ quantifies their relative contributions, with $\tan \delta > 1$ and $\tan \delta < 1$ indicating largely dissipative and elastic dynamics, respectively. As such, Figs 2A and S1 signify that ApoI digestion increases elastic-like contributions. This result is rather surprising given that ApoI cuts the full-length plasmids into many smaller fragments, reducing the polymer overlap and entanglement density. At first glance, we expect this process to reduce elastic contributions to the stress response. However, because the DNA polymers start out as rings, enzymatic digestion first linearizes them, leading to linear chains of the same length as the rings, followed by more and more fragmentation of the linear chains as digestion proceeds. Because linear polymers form entanglements that are longer-lived and fundamentally different than the steric interactions between highly-overlapping rings of equal length (*6–9, 12, 17, 31–34*), and ring-linear threadings are more persistent than ring-ring threadings, enzymatic linearization may underlie the increase in $G'$ and decrease in $\tan \delta(\omega)$.

To investigate these potential mechanisms, and elucidate the rheological effects of linearization and subsequent fragmentation, we evaluate the time-dependent topological conversion (Fig 1C) and bulk rheology (Fig 2B,C) over the course of the 6-hour ApoI digestion. Our time-resolved gel electrophoresis measurements show that DNA digestion proceeds steadily over the course of several hours (Fig 1C), suggesting that we may likewise expect $G'(t_a)$ to initially steadily increase as DNA molecules are linearized, followed by a steadily decreasing $G'(t_a)$ as fragmentation ensues. Contrary to these expectations, our time-resolved bulk oscillatory measurements performed at ApoI digestion times $t_a \simeq 0 - 360$ mins, show an abrupt transition from a state with a relatively $t_a$-independent elastic modulus value of $G'_i \simeq 3$ mPa to a state characterized by an order of magnitude higher $G'$ value (Fig 2B). The time-varying viscous modulus $G''(t_a)$ displays a similar discrete state transition but with a smaller difference between the two states (Fig 2B). The transition to increased elasticity can also be clearly seen by evaluating $\tan \delta(t_a)$, which drops sharply from $\langle \tan \delta \rangle > 1$ (dissipative) to $\langle \tan \delta \rangle < 1$ (elastic-like) at $t_{a,s1} \simeq 160$ min (Fig 2C). Of note, this transition approximately coincides with the time at which complete digestion is reached, denoted by the green arrow in Fig 1C.

***In situ topological conversion of DNA tunes the bulk rheology of DNA-dextran composites with non-monotonic dependence on $\phi_{DNA}$.*** We now aim to understand if the distinct rheological state-switching we observe in DNA solutions is transferrable to composite systems of DNA and synthetic polymers (e.g., dextran); and, if so, to what extent the non-equilibrium rheological properties may be tuned by the inclusion of such polymers.

We first examine the steady-state rheology of $11c^*$ DNA-dextran composites with varying volume fractions of DNA, $\phi_{DNA} = 0.25, 0.5, 0.75$, after 6 hours in the presence or absence of ApoI (Fig 3A-C), analogous to Fig 2A for the pure DNA solution ($\phi_{DNA} = 1$). In contrast to $\phi_{DNA} = 1$, we find that ApoI digestion actually reduces $G'(\omega)$ for all DNA-dextran composites across nearly the entire frequency range, indicating that DNA digestion reduces rather than enhances the bulk elasticity.

Another notable distinction between the pure DNA solution and the composites, is the presence of a crossover frequency $\omega_c$ at which $G''(\omega)$ becomes larger than $G'(\omega)$, indicated by arrows in Fig 3A-C. This crossover frequency is a measure of the fastest relaxation timescale of the system, i.e., $\tau_f \simeq 2\pi/\omega_c$. For entangled linear polymer systems, $\tau_f$ is predicted to be the entanglement time $\tau_e$, which is the timescale over which an entangled polymer reaches the edge of the entanglement tube and thus 'feels' the constraints of entanglements (*35*). As such, smaller and larger $\tau_e$ values (larger and smaller $\omega_c$) generally equate to higher and lower entanglement densities, respectively.

We quantify $\omega_c$ for each DNA-dextran composite as the frequency at which $\tan \delta(\omega) = 1$ (Fig 3D), and compare to extrapolated lower-bound estimates of $\omega_c$ for $\phi_{DNA} = 1$. Figure 3D shows the dependence of $\omega_c$ on $\phi_{DNA}$, with and without ApoI digestion. The solid and dashed lines denote, respectively, the predicted scaling of $\omega_c \sim \phi^{2.5}$ for entangled linear polymers (*36*), as well as the empirical scaling $\omega_c \sim \phi_{DNA}^{5.3}$ previously reported for composites of entangled linear DNA and dextran (*25*). After digestion, $\omega_c$ scales approximately as $\phi^{2.5}$ for $\phi_{DNA} \leq 0.5$ but transitions to scaling closer to $\phi^{5.3}$ for $\phi_{DNA} > 0.5$. This steeper scaling was previously shown to arise from dextran-mediated stretching and bundling of linear DNA in DNA-dextran composites, which caused the DNA network to behave as if comprising semiflexible rather than flexible polymers (*25, 37, 38*). We note that, without digestion, there is no discernible scaling of $\omega_c$ with $\phi_{DNA}$.

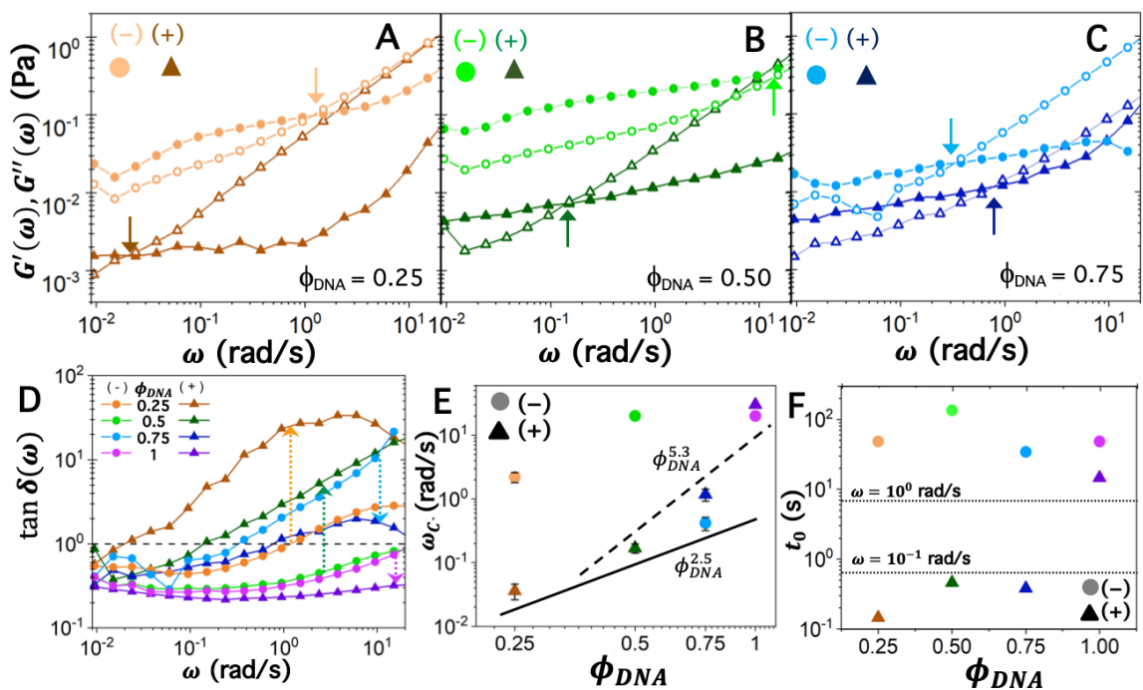

**Figure 3. DNA-dextran composites exhibit signatures of both increased and decreased viscoelasticity following enzymatic fragmentation of circular DNA.** (**A-C**) Frequency-dependent linear viscoelastic moduli, $G'(\omega)$ (filled symbols) and $G''(\omega)$ (open symbols), of $11c^*$ composites of DNA and dextran at varying DNA fractions: $\phi_{DNA} = 0.25$ (**A**, orange), $\phi_{DNA} = 0.5$ (**B**, green), and $\phi_{DNA} = 0.75$ (**C**, blue), measured after 6 hours of either ApoI digestion ((+), triangles, darker hues) or under identical conditions but without ApoI ((-), circles, lighter hues). All composites exhibit crossover frequencies $\omega_c$, indicated by color-coded arrows, that depend on the digestion state and $\phi_{DNA}$. (**D**) Frequency-dependent loss tangents $\tan \delta(\omega)$ derived from the data shown in (A-C) and Fig 2A for all composites and digestion states. Color-coded dotted arrows point from the (-) to (+) condition for each $\phi_{DNA}$. (**E**) Crossover frequency $\omega_c$, determined by evaluating the frequency at which $\tan \delta(\omega) = 1$ (denoted by the dashed horizontal line in D) for $\phi_{DNA} = 0.25$ (orange), 0.5 (green), 0.75 (blue), and 1 (purple), measured after 6 hours with ApoI ((+), dark shades, triangles) and without ApoI (-, light shades, circles). Note that $\omega_c$ values for $\phi_{DNA} = 1$ are lower bound approximates as described in the text. Solid and dashed lines correspond to scaling relations $\omega_c \sim \phi^{2.5}$ and $\omega_c \sim \phi^{5.3}$ described in the text. (**F**) Structural relaxation times $t_0 \simeq 3\pi\eta_0 R_G^3/k_B T$ versus $\phi_{DNA}$ for the initial (-, circles) and final (+, triangles) states comprising either full-length rings (-, $l_0 \simeq 3667$ nm) or linear fragments (+, $\langle l_f \rangle \simeq 854$ nm). Dashed horizontal lines indicate timescales equivalent to the frequencies listed.

Another important timescale that may influence how molecular-level topological conversion maps to bulk rheological properties in the different composites is the structural relaxation time $t_0$ of the digested and undigested DNA, which is an estimate of how quickly the DNA conformation becomes decorrelated from its initial conformation (i.e., it 'forgets' its previous state). For our topologically-active systems, $t_0$ indicates how long it may take for an enzymatic cleavage event to result in a new structural state of the cleaved DNA, and thus, a new rheological state of the system. We consider $t_0$ as the time for DNA to diffuse a distance comparable to its radius of gyration, i.e., $t_0 \simeq R_G^2/2D \simeq 3\pi\eta_0 R_G^3/k_B T$ where $D$ is the DNA diffusion coefficient and $\eta_0$ is the zero-shear viscosity of the network. We estimate a lower-bound for $\eta_0$ of the digested and undigested states of each composite from the low-$\omega$ limit of our measured bulk complex viscosities (SI Fig S3). Using values of $R_G \simeq 196$ nm and $R_G \simeq 69$ nm for the initial full-length ring DNA ($l_0 \simeq 3667$ nm) and the final average DNA fragment ($\langle l_f \rangle \simeq 854$ nm) along with $\eta_0$ values plotted in SI Fig S3, we determine structural relaxation times of ~0.15 s to 135 s (Fig 3F). In comparison, the fastest and slowest frequency measurements shown in Figs 2,3 take ~0.3 s and ~700 s to complete. As $t_0$ is comparable to the data acquisition times (Fig 3F), we do not expect there to be significant 'delay' in the enzymatic digestion reflected in the rheology. Fig 3F also shows that DNA in the digested state relaxes much more slowly in the $\phi_{DNA} = 1$ solution compared to the composites, while in the undigested state, $\phi_{DNA} = 0.5$ exhibits the slowest structural relaxation. This effect can also be seen in Fig 3E which shows that the $\phi_{DNA} = 0.5$ composite has the highest $\omega_c$ value among the systems, suggesting that free DNA diffusion is suppressed the most.

To estimate $c_e$ for the fully digested composites, we recall that ApoI digestion results in fragments with minimum, maximum and average lengths of $l_{f,min} \simeq 64$ bp, $l_{f,max} \simeq 3160$ bp, and $\langle l_f \rangle \simeq 854$ bp, respectively (*41*). Combining the relations $c^* \sim l R_G^{-3}$ and $R_G \sim l^\nu$, where $\nu \simeq 0.588$ is the Flory exponent for flexible polymers in good solvent conditions (which we have here) (*21, 35, 40, 42–44*), yields $c_L^*(l) \sim l^{1-3\nu}$, from which we can derive an entanglement concentration expression $c_e(l_f) \simeq 6 \times (\langle l_f \rangle/l_i)^{1-3\nu} \times 0.25 c_R^*(l_i)$. Using $c_R^*(l_i) \simeq 480$ μg/mL, $l_i = 11$ kbp and $l_f \approx \langle l_f \rangle$, we compute $c_e(l_f) \simeq 5.1$ mg/ml. It follows that $\phi_{DNA} = 1, 0.75, 0.5$ and $0.25$ composites correspond to ~$1.04 c_e(l_f), 0.78 c_e(l_f), 0.52 c_e(l_f)$ and $0.26 c_e(l_f)$. Alternatively, we can compute the expected critical entanglement molecular weight $M_c$ or length $l_c$, which is the polymer length, in units of monomers ($M_c$) and physical length ($l_c$), above which the system exhibits entanglement dynamics at a given concentration. Provided that $c_e \simeq 720$ μg/mL for 11 kbp DNA, we can infer that $l_c \simeq 11$ kbp at $c \simeq 720$ μg/mL, and use the scaling relation $l_c \sim l_e \sim c^{1/(1-3\nu)}$ (*35, 45*) to compute $l_c$ values of ~825 bp, ~1.2 kbp, ~2 kbp and ~5 kbp for $\phi_{DNA} = 1, 0.75, 0.5$ and $0.25$.

As $c > c_e(l_f)$ and $l_c < \langle l_f \rangle$ for $\phi_{DNA} = 1$, we can expect the solution to remain entangled even when fully digested, which may explain why we observe an increase in $G'$ and decrease in $\tan\delta$ following digestion (Fig 2). An additional contribution to the increased $G'$ and reduced $\tan\delta$ may be the conversion of rings to linear chains which is expected to increase entanglements and threadings. Conversely, the concentrations of $\phi_{DNA} < 1$ composites are all less than $c_e(l_f)$ and their $l_c$ values are likewise higher than $\langle l_f \rangle$, so we expect minimal entanglements after digestion, and, therefore, lower $G'$ and higher $\tan\delta$ values as compared to the undigested composites. This trend is indeed what we observe for $\phi_{DNA} = 0.25$ and $0.5$ composites, but is at odds with the higher $\omega_c$ and lower $\tan\delta$ that we measure for the digested $\phi_{DNA} = 0.75$ composite.

To reconcile this seeming contradiction for $\phi_{DNA} = 0.75$, we note that the scaling arguments described above do not consider the effects of dextran crowding, such as the previously reported entropic stretching and bundling of linear DNA in DNA-dextran composites (25), which may either promote connectivity at high $\phi_{DNA}$ or hinder connectivity as $\phi_{DNA}$ is reduced. Specifically, these previous studies showed that $\phi_{DNA} = 0.75$ composites exhibited a ~2× higher plateau modulus and ~4× slower DNA diffusion compared to their $\phi_{DNA} = 1$ counterparts due to bundling and increased self-association of the linear DNA, which locally increased the effective entanglement density and polymer stiffness while maintaining network connectivity. We conjecture that the same phenomenon is occurring in the digested $\phi_{DNA} = 0.75$ composite, whereas for lower $\phi_{DNA}$ composites, bundling comes at the cost of destroying the DNA network connectivity necessary to elastically resist stresses.

***DNA fragmentation elicits abrupt and tunable rheological state-switching of DNA-dextran composites.*** Armed with a mechanistic understanding of the $\phi_{DNA}$-dependent initial and final rheological states of the composites (Fig 3), we turn to discovering how their rheological properties evolve during ApoI digestion (Fig 4).

As shown in Fig 4A,B, similar to $\phi_{DNA} = 1$, each composite displays an abrupt transition from an initial state, described by a time-averaged storage modulus $G'_i$, to a state characterized by higher, nearly constant $G'(t_a)$ values, the average of which we denote as $G'_h$. However, unlike $\phi_{DNA} = 1$, after lifetime $\Delta t_h$ in the $G'_h$ state, each composite undergoes a subsequent, similarly sharp drop to a final state characterized by a lower average modulus $G'_l$ which is similar in magnitude to the initial state $G'_i$. This rheological state-switching is also apparent in the corresponding $G''(t_a)$, $\tan \delta(t_a)$, and $\eta^*(t_a)$ curves for each composite (SI Fig S2). Importantly, this rather surprising non-equilibrium rheological behavior, cannot be captured or predicted by solely evaluating the initial and final rheological states (Fig 3), as they lack all information about the intermediate high-elasticity $G'_h$ state.

While the two-state behavior is surprisingly robust across the entire composite phase space, the values of $G'_i$, $G'_h$, $G'_l$, and $\Delta t_h$, as well as the times at which the initial jump and subsequent drop occur, $\tau_{s1}$ and $\tau_{s2}$, depicted in Fig 4B, are tuned by $\phi_{DNA}$ (Fig 4C-E). To understand the dependence on $\phi_{DNA}$ we first observe that $G'_h$ exhibits a power-law dependence on $\phi_{DNA}$ that is remarkably similar to the predicted scaling $G^0 \sim c^{2.25}$ that relates the plateau modulus $G^0$ to concentration $c$ for entangled linear polymer solutions (35, 45). This scaling is derived from the expression $G^0 = \left(\frac{c}{N}\right) k_B T \left(\frac{c}{c^*}\right)^{1.25}$ (45), which, for DNA, equates to $G^0 = (3.8c^*/L)(c/c^*)^{2.25}$ where $L$ is the DNA length in units of basepairs, $c^*$ is in µg/mL, and $G^0$ is in Pa. Using input parameters for the full-length linear DNA, i.e., $N = 11$ kbp, $c^* \simeq 120 \frac{\mu g}{mL}$, and $c/c^* \simeq 44\phi_{DNA}$ (recall that $c^*_R \simeq 4c^*_L$), we find remarkable agreement between the theoretical predictions for $G^0$ and our measured $G'_h$ values. The same calculation using input parameters for the rings yield ~5x lower $G^0$ values. This result indicates that the elastic storage in the high-elasticity state is dictated primarily by entanglements between the long linear chains in the composites.

The modulus in the subsequent low-elasticity state $G'_l$ also scales with $\phi_{DNA}$ but with a much weaker dependence. As we describe above, if we neglect any effects of dextran, the $\phi_{DNA} < 1$ composites are not expected to exhibit entanglement dynamics following complete digestion. In this case, the scaling of $G'_l(\phi_{DNA})$ should follow the predicted concentration dependence for $G'$ for semidilute unentangled polymer solutions, $G' \sim c$. Fig 4D shows good agreement with this predicted scaling $G'_i \sim \phi_{DNA}$, confirming that indeed the composite rheology is dominated by the dynamics of unentangled polymers.

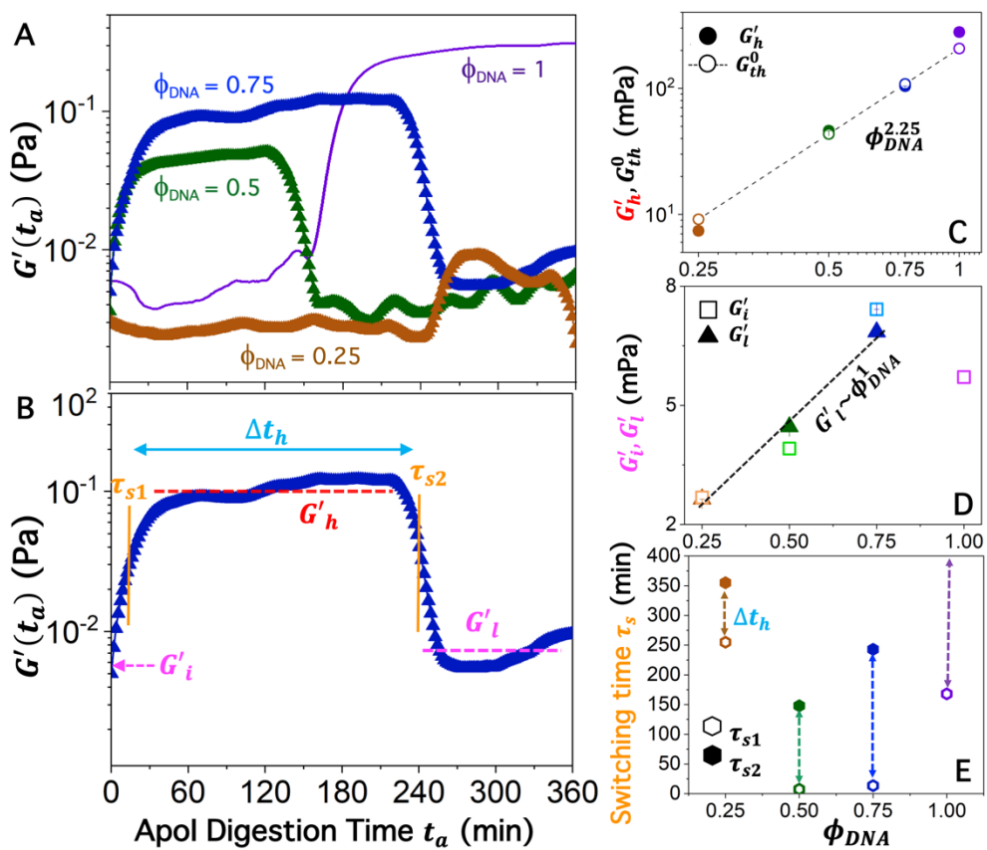

**Figure 4. DNA digestion induces abrupt switching between discrete viscoelastic states of DNA-dextran composites.** Elastic modulus $G'(t_a)$ versus digestion time for composites with $\phi_{DNA} = 0.25$ (orange), 0.5 (green), and 0.75 (blue), compared to $\phi_{DNA} = 1$ (purple line). All composites exhibit sharp transitions between distinct viscoelastic states defined by metrics shown in **(B)** and plotted in (C-E). **(B)** $G'(t_a)$ for $\phi_{DNA} = 0.75$ are shown as sample data to depict how metrics are defined and computed. **(C)** Average elastic modulus in the 'high-elasticity' state, $G'_h$ (filled circles, red dashed line in (B)) and theoretically predicted values for the elastic plateau modulus $G^0_{th}$ (open symbols) versus $\phi_{DNA}$. The dashed line connecting theoretical values follows the predicted power-law $G^0_{th} \sim \phi_{DNA}^{2.25}$, in good agreement with experimental $G'_h$ values. **(D)** Average elastic moduli in the initial ($G'_i$, open squares) and final 'low-elasticity' ($G'_l$, filled triangles) states in which $G'(t_a)$ exhibits minimal $t_a$-dependence. Dashed scaling line denotes the theoretically predicted linear scaling $G' \sim \phi$ for semidilute unentangled polymers, which aligns with $G'_l$ scaling. $G'_i$ values follow a non-monotonic dependence on $\phi_{DNA}$. Data points in C,D represent the mean and standard error across each $t_a$-independent region. **(E)** The times at which composites jump to high-elasticity states ($\tau_{s1}$, open) and then drop to low-elasticity states ($\tau_{s2}$, filled), depicted by orange vertical lines in (B), are defined as the time at which $G'(t_a)$ is halfway between the corresponding initial and final state values. Vertical arrows point from $\tau_{s1}$ to $\tau_{s2}$ with the arrow length indicating the relative lifetime $\Delta t_h$ of the $G'_h$ state.

The more complicated scaling with $\phi_{DNA}$ manifests for the initial low-elasticity state modulus $G'_i$. In this case, we observe, once again, a non-monotonic dependence of $G'_i$ on $\phi_{DNA}$ with a maximum at $\phi_{DNA} = 0.75$. As such, composites in this state cannot simply be modeled as comprising unentangled polymers. Rather, this finding bolsters our understanding that dextran serves to bundle the DNA polymers such that their stiffness and effective local concentration is higher than for a pure DNA solution at the same concentration, leading to a higher modulus for $\phi_{DNA} = 0.75$ as compared to $\phi_{DNA} = 1$ (*25*). However, bundling in the $\phi_{DNA} \leq 0.5$ composites comes at the cost of connectivity which drops the modulus to below expected values, as seen by comparing the scaling of $G'_l$ for $\phi_{DNA} \leq 0.5$ to the predicted linear scaling $G'_l \sim \phi_{DNA}$.

Finally, and perhaps most surprising, the initial state transitions for the $\phi_{DNA} = 0.5$ and $\phi_{DNA} = 0.75$ composites occur nearly instantaneously, with no discernable $G'_i$ plateau (Fig 4A,E), in contrast to the $\phi_{DNA} = 1$ transition time of $\tau_{s1} \simeq 168$ min. At the same time, the initial jump for $\phi_{DNA} = 0.25$ occurs much later than the other systems, with $\tau_{s1} \simeq 255$ min, and the corresponding elastic lifetime $\Delta t_h$ is ~29% and 56% shorter than $\phi_{DNA} = 0.5$ and $\phi_{DNA} = 0.75$, respectively (Fig 4E). These $\phi_{DNA}$-dependent transition times and elastic lifetimes, may indicate dextran-mediated alterations to the efficiency and/or rate of DNA digestion, as well as changes to the structural relaxation times of the polymers.

## Discussion

Two counterintuitive features of our data that require further discussion are the: [1] rapid onset of the elastic state in $\phi_{DNA} = 0.5$ and $\phi_{DNA} = 0.75$ composites as compared to the pure DNA solution (Figs 4E, 5E), and [2] universal sharpness and discrete nature of the state transitions. [1] The rapid onset of the $G'_h$ state may indicate that dextran accelerates digestion by facilitating DNA-ApoI interactions via depletion interactions, a common occurrence in crowded systems (*46–49*). [2] The sharpness of the transitions may be suggestive of cooperative mechanisms required to elicit bulk changes to the system from molecular-level activities. As described in the Introduction, cooperativity frameworks have been used to describe glass transitions in polymeric materials as well as dynamics of threaded rings (*15*, *18*, *19*), suggesting potential applicability here. We unpack these hypotheses in turn below.

***DNA-dextran interactions enhance DNA overlap and slow dissipative modes with minimal impact on topological activity.*** To quantify the kinetics of topological conversion for each composite, we perform time-resolved gel electrophoresis (see Methods). Figure 5A-D shows that, counter to depletion-driven acceleration conjectured above, the time to complete digestion $t_d$ (dotted lines) modestly increases for $\phi_{DNA} = 0.5$ and $\phi_{DNA} = 0.75$ compared to $\phi_{DNA} = 1$. Figure 5 also reveals that the elastic onset for these composites coincides roughly with the time at which DNA rings have been converted entirely to linear topology (large-dashed lines), in contrast to $\phi_{DNA} = 1$ in which it does not occur until digestion is complete.

Moreover, the time of the subsequent drop (i.e., $\tau_{s2}$) for $\phi_{DNA} = 0.75$ and $\phi_{DNA} = 0.5$ align with the times at which the largest fragment drops to $l_{f,max} \approx 4$ kbp and $l_{f,max} \approx 6.5$ kbp, respectively, as estimated from gel electrophoresis. Considering the critical entanglement lengths $l_c \approx 2$ kbp and ~1.2 kbp for $\phi_{DNA} = 0.5$ and $\phi_{DNA} = 0.75$, we can estimate the critical maximum polymer length required to maintain bulk elasticity to be $l_h \approx 3.3 l_c$. Using similar scaling arguments also provides a critical concentration of $c_h \approx 2.5 c_e$ to maintain a high-elasticity state.

These critical parameters equate to $l_h \approx 2.7$ kbp and $c_h \approx 4.8$ mg/ml for the fully digested $\phi_{DNA} = 1$ solution, which are below $l_{f,max} \approx 3.2$ kbp and $c_{\phi=1} \approx 5.3$ mg/ml, respectively;

supporting the persistence of the elastic state for $\phi_{DNA} = 1$ well after digestion completes. We can likewise estimate $l_h \approx 3.3 l_c \approx 16$ kbp and $c_h \approx 1.8$ mg/ml for $\phi_{DNA} = 0.25$, which are larger than even the undigested $\phi_{DNA} = 0.25$ values of $l_i \approx 11$ kbp and $c_{\phi=0.25} \approx 1.3$ mg/ml. This argument suggests that digestion should have minimal impact on the rheology of $\phi_{DNA} = 0.25$, as evidenced by the shorter elastic lifetime $\Delta t_h$ and delayed elastic onset $\tau_1$ compared to the other composites (Fig 5E).

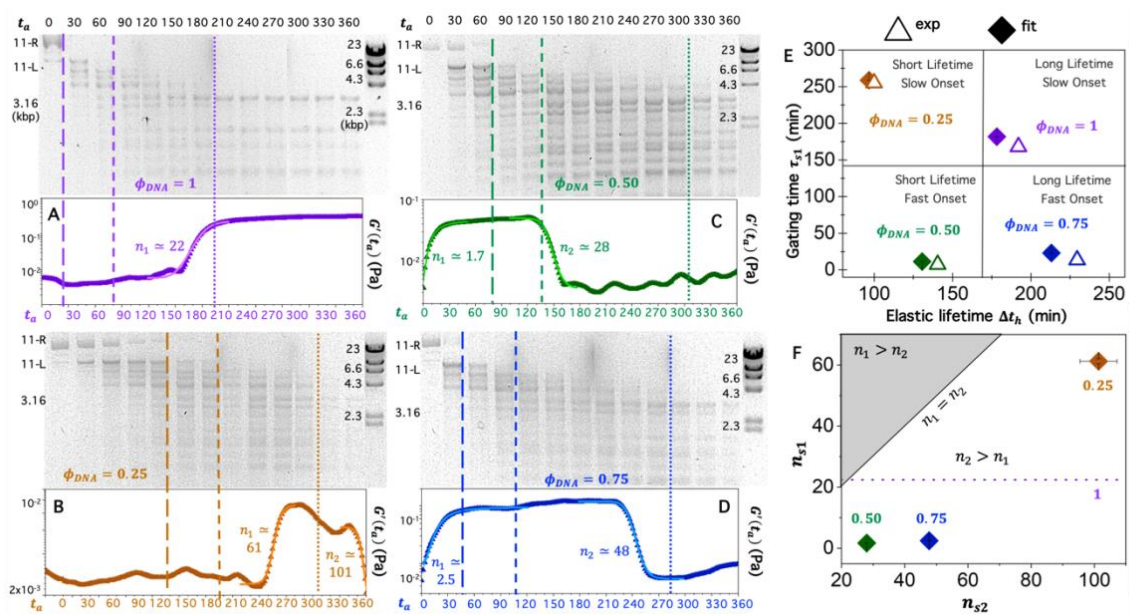

**Figure 5. Elastic gating times and lifetimes do not mirror enzymatic digestion kinetics and are tuned by $\phi_{DNA}$.** Time-resolved gel electrophoresis of DNA undergoing ApoI digestion in composites with $\phi_{DNA} = 1$ (**A**), 0.25 (**B**), 0.50 (**C**), and 0.75 (**D**). Each lane shows the topological state of the DNA at a given time $t_a$, listed in mins. Far-right lanes are the $\lambda$-HindIII molecular weight marker with the size of linear DNA corresponding to select bands listed in kilobasepairs (kbp). Labels to the left denote the bands that correspond to 11 kbp ring DNA (11-R), 11 kbp linear DNA (11-L), and the largest fragment length (3.16 kbp). Vertical lines indicate times at which: ring DNA is completely converted to linear (large-dashed), all full-length linear DNA has been cleaved (small-dashed), and DNA has been completely digested (dotted). $G'(t_a)$ curves below each gel correlate the bulk rheological state with the topological state of DNA. Solid lines are fits to effective Hill functions: $G'(t_a) = (G'_{h/l} - G'_{i/h})/(1 + \left(\frac{\tau_{s1/s2}}{t_a}\right)^{n_{1/2}}) + G'_{i/h}$. (**E**) Elastic gating time $\tau_{s1}$ versus lifetime $\Delta t_h$ determined from the data (open triangles) and fits (filled diamonds). $\Delta t_h$ values for $\phi_{DNA} = 1$ are lower bounds as $G'_l$ is not reached within the measurement time. Horizontal and vertical lines divide the data into different combinations of short/long lifetime and fast/slow onset of elasticity. (**F**) Cooperativity parameters associated with the initial jump $n_{s1}$ and subsequent drop $n_{s2}$. $n_{s1}$ for $\phi_{DNA} = 1$ is denoted by the purple dotted line as it has no corresponding $n_{s2}$. Diagonal $n_{s1} = n_{s2}$ line divides the plot into regions in which $n_{s1} > n_{s2}$ (grey) and $n_{s1} < n_{s2}$.

We now turn to understanding the sharpness of the transitions between distinct rheological states that do not mirror the topological digestion kinetics. As seen in Fig 5, digestion proceeds steadily over the course of the experimental time and the structural relaxations of the molecules are sufficiently fast to expect steadily changing bulk rheological properties as the DNA topologies and lengths are changing (Fig 3F). However, this is not what we find. Instead, we observe rapid sigmoidal-like transitions between the different states, a hallmark of many two-state systems in which the steepness of the curve indicates more or less cooperativity. This concept is generalizable and used to describe many disparate systems and processes (e.g., *15*, *50–56*). For example, cooperativity is a hallmark of DNA melting, micelle formation, protein folding, DNA stretching and nucleic acid hairpin unfolding (*57–63*). Cooperativity has also been suggested to play a principal role in the dynamics of concentrated ring polymers and ring-linear blends (*64*); and to drive glass transitions in amorphous materials in which the motion of one constituent in the system requires the cooperative motion of many others (*18*, *50*, *65–69*). In all of these cases, molecular-level processes give rise to larger-scale structural changes, but require cooperativity between the molecular-level components to allow the large-scale structural change to take place. Likewise, we suggest that the phenomenon we observe is due to cooperativity between molecular-level constituents giving rise to the alterations in bulk rheological properties.

Cooperative transitions between two states often exhibit sigmoidal functional dependence on the independent variable (e.g., temperature for DNA melting and protein folding, force for DNA stretching and hairpin unfolding, amphiphile concentration for micelle formation). The sharper or steeper the sigmoid, the more cooperative the process is. A widely used function to describe such two state transitions is the Hill function, often used to describe the fraction of receptors (e.g., macromolecules) bound by ligands (e.g., enzymes) as a function of ligand concentration (*65*, *66*). The Hill function has an explicit cooperativity term $n$ that quantifies the steepness of the sigmoid as a read-out for cooperativity, with $n > 1$, $n < 1$ and $n = 1$ indicative of cooperativity, antagonism, and traditional Michaelis-Menten (MM) kinetics, respectively (*65*, *66*, *70*, *71*). Accordingly, we derive a sigmoidal model analogous to the Hill function to fit our $G'(t_a)$ data, to confirm that a cooperative model can describe our results and quantify the degree of cooperativity. In our model, we treat $G'$ and $t_a$ as synonymous with the bound receptor fraction and ligand concentration, respectively, leading to the function $G'(t_a) = [(G'_2 - G'_1)/(1 + \left(\frac{\tau_s}{t_a}\right)^n)] + G'_1$ where $G'_1$ and $G'_2$ are the elastic moduli of the first and second states and $\tau_s$ is the characteristic switching time. $G'_1$, $G'_2$ and $\tau_s$ values from each fit are measures of $G'_i$, $G'_h$ and $\tau_{s1}$ for the first state switch and $G'_h$, $G'_l$ and $\tau_{s2}$ for the second, providing an alternative method for determining these values and verifying the goodness of the fits (SI Table S1). We find that all state transitions are well-fit to the model with $\phi_{DNA}$-dependent fitting parameters (Fig 5). We also find that all composites exhibit cooperative state switching ($n_1$, $n_2 > 1$) with the second transition being notably more cooperative than the first (i.e., $n_2 > n_1$) (Fig 5F, Table S1).

To understand the physical picture underlying the signatures of cooperativity, we turn to previous works that use the general concept of cooperativity or correlations between molecular components to describe bulk glass transitions in a wide range of systems (*18*, *19*, *50*, *68*, *69*, *72*, *73*). The physical cooperativity picture is one of discrete clusters or rafts of slow-moving particles that grow over time until they reach a critical size or connectivity at which point the system motion becomes dynamically arrested or glass-like. Leading up to the glass transition, these discrete 'clusters' of slow-moving, correlated molecules form amidst a bath of faster-moving constituents, resulting in heterogeneous dynamics. Previous studies have noted the universality of cooperativity in glass transitions, with the steepness of the transition depending only on the number of correlated molecules and being insensitive to the type of molecule or inter-molecular interactions (e.g., *50*, *56*). Moreover, the size of the clusters has been shown to directly relate to the emergent elasticity (*19*).

In light of these works, we argue that the discrete state-switching in our systems arises from formation, growth and eventual percolation of cooperative clusters of slow-moving entangled DNA (Fig 6A). Analogous to continuously changing temperature in glass transitions, enzymatic digestion proceeds continuously over time. While the topologies and sizes of the molecules are changing steadily and continuously (Fig 5A-D), 'slow' cooperative clusters form, grow and connect until they percolate and span the system. At the macroscopic scale, we expect this cooperativity to lead to periods of nearly constant low elasticity, as the entangled clusters grow and form connections with their neighbors, followed by an abrupt transition to an elastic-like response once the cluster connectivity reaches percolation. Within this framework, the elastic transition/onset time $\tau_{s,1}$ is analogous to the glass transition temperature in glass formers. In the broader context of cooperativity driving discrete state transitions, $\tau_{s,1}$ can be thought of in analogy to the, e.g., DNA melting temperature (*51*, *56*), critical ring length and density for glassy dynamics in ring polymer melts (*15*), critical amphiphile concentration for micelle formation (*52*, *53*), and critical force to unfold nucleic acid hairpins (*54*, *55*).

We can qualitatively understand the second state transition, from high to low elasticity, which exhibits more positive cooperativity than the first switch (Fig 5F), as the reverse of the general physical model for elastic onset. Specifically, in the $G'_h$ state, linear DNA chains forming a fully-connected entangled network, begin to lose connections as they are enzymatically cleaved into shorter fragments, eventually losing enough entanglements to no longer percolate throughout the system (Fig 6A), at which point the system can no longer support bulk stresses and thus undergoes a sharp drop in the elastic response. The enhanced sharpness of this transition compared to the first may arise from the reduced complexity and fewer competing factors. Namely, we attribute the first transition primarily to the linearization of ring DNA, which increases coil overlap and thus entanglements. Yet, as rings are depleted from the system, threading events, which themselves suppress dissipation and thus contribute cooperatively to the transition, are likewise reduced. At the same time, competition between linearization, that swells the polymer coils, and fragmentation that reduces entanglement density, also compete to reduce the cooperativity of the first transition compared to the second.

We acknowledge that our system is indeed different than typical glass-forming systems, and the transitions we report are not traditional glass transitions. Namely, we do not reach complete dynamic arrest, and the state-switching we observe is a function of time rather than temperature or polymer concentration, as in the cases of glass-formers and glassy dynamics of threaded rings, respectively. Nevertheless, as described above, previous works have highlighted the generality of the physical picture of growing cooperative clusters as describing glass-like transitions in a wide range of disparate systems, noting that the only relevant parameters are the number of correlated constituents (whatever they may be) and the size of the clusters they comprise (relative to the system size). For these reasons, we adopt this physical model to rationalize the discrete state-switching phenomena that we observe.

***Heterogeneous DNA dynamics give rise to bulk rheological state-switching that can be programmed at the molecular level.*** As described in the Introduction, a universal feature of systems near glass transitions has been shown to be non-Gaussian van Hove distributions of the constituents (*20*) that have large-displacement tails that extend beyond the Gaussian profile and may also exhibit higher probabilities of near-zero displacements. The former is due to the fast-moving population while the latter is a signature of the slow-moving clusters. To confirm that our systems indeed exhibit similar heterogenous dynamics, we tracked single fluorescent-labeled DNA molecules within the composites over the same 6-hr time course of our rheology and electrophoresis experiments. Figure 6B-G shows the van Hove distributions for key time points during the digestion of the $\phi_{DNA} = 0.50$ composite, where distinct non-Gaussian tails can be seen during the transitions (Fig 6B-D). In contrast, in the

high-elasticity state, in which we expect percolation and minimal heterogeneity, the distributions are primarily Gaussian (Fig 6E,F).

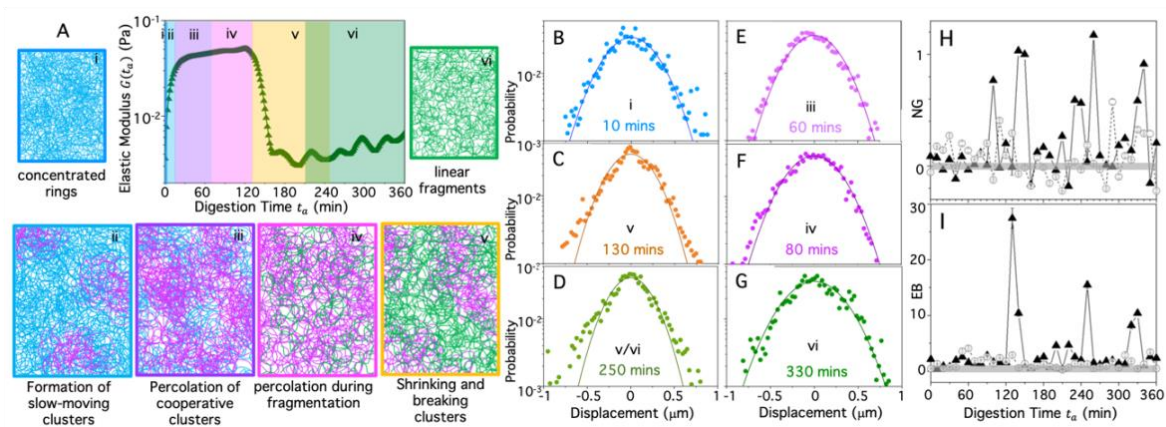

**Figure 6: Cooperative molecular mechanisms underlying glass transitions can explain the rheological state-switching of DNA-dextran composites**. (**A**) Cartoon depiction of cooperative process underlying state-switching of concentrated DNA rings (blue), that are enzymatically converted to linear chains (magenta) which are then chopped into smaller linear fragments (green). The qualitative description is inspired by the cooperative molecular mechanisms that underlies bulk glass transitions, as discussed in the text. (Left to Right) Some of the initially ring DNA (**i**) gets cleaved into linear chains that form cooperative slow-moving clusters (**ii**) that grow, connect, and eventually percolate (**iii**). As the linear chains are fragmented, the initially percolated clusters (**iv**) begin to shrink, disconnect (**v**), and eventually disappear (**vi**). (**B-G**) Sample van Hove distributions for $\phi_{DNA} = 0.5$ at times $t_a = 10$ (**B**), 130 (**C**), 250 (**D**), 60 (**E**), 80 (**F**), and 330 (**G**) mins show non-Gaussian displacement distributions that are most apparent during the state transitions ($t_a = 10, 130, 250$). (**H**) The average non-Gaussianity parameter $NG$ and ergodicity breaking parameter $EB$ computed for $\phi_{DNA} = 0.5$ with (solid triangles) and without (open circles) ApoI, plotted versus $t_a$, show that $NG$ and $EB$ are generally larger for the topologically-active composite compared to the control, and exhibit larger than average peaks at the transition points.

To quantify the extent to which the displacements deviate from Gaussianity we compute the non-Gaussianity parameter $NG$ and ergodicity breaking parameter $EB$, as described in Methods (*74, 75*), over the course of the 6-hr digestion. For spatiotemporally homogenous, Gaussian dynamics, $NG \approx EB \approx 0$, and the larger the value of $|NG|$ and $|EB|$ the more non-Gaussian and non-ergodic the transport is. We first evaluate $\langle NG \rangle$ and $\langle EB \rangle$, averaged over all 37 measurements over the course of the 6-hr digestion, measuring $\langle NG \rangle = 0.23 \pm 0.06$ and $\langle NG \rangle = 0.17 \pm 0.06$ for $\phi_{DNA} = 0.50$ and $\phi_{DNA} = 1$ compared to $\langle NG \rangle = 0.07 \pm 0.03$ for the control $\phi_{DNA} = 0.5$ case without ApoI. Similarly, we find $\langle EB \rangle = 3.5 \pm 0.9$ and $\langle EB \rangle = 3.6 \pm 1.0$ for $\phi_{DNA} = 0.50$ and $\phi_{DNA} = 1$ compared to $\langle EB \rangle = 1.4 \pm 0.2$. These data show that composites undergoing digestion clearly deviate from Gaussianity more strongly than the control composite. Moreover, the time-dependence of both parameters, shown in Fig 6H,I, exhibits substantially amplified non-Gaussianity near the transition times $\tau_1$ and $\tau_2$. Amplified $NG$ and $EB$ values can also be seen at later times during digestion when we expect the percolation to be broken and for the slow-moving clusters to become progressively smaller, leading to heterogeneous transport characteristics. Note that in the control case, not only are $NG$ and $EB$ smaller, on average, but there is also much less variation of the values over time. These data confirm that the topologically-active composites exhibit universal signatures of glass transitions, further supporting our proposed mechanistic description of the phenomena, as depicted in Fig 6A.

In conclusion, our results demonstrate that molecular-level topological conversion of circular DNA can be harnessed to engineer bio-synthetic composite materials that exhibit robust and tunable non-equilibrium rheological properties. We discover that steady enzymatic linearization and fragmentation of DNA rings leads to sharp switching between discrete rheological states of DNA-dextran composites, characterized by more or less elasticity. We rationalize the sharp transitions, which can be described by sigmoidal Hill-like functions, as arising from cooperative percolation and dissolution of slow-moving DNA clusters, analogous to glass transitions, which manifest as bulk states that largely store or dissipate energy. Moreover, we show that the sharpness and strength of the state-switching, as well as the gating time and lifetime of the elastic-like state can be precisely tuned over orders of magnitude by varying the relative fractions of the topologically-active substrate (DNA) and synthetic polymer counterpart (dextran) of the composite.

We anticipate that our approach of leveraging topological operations of molecular constituents to confer tunable time-varying bulk rheology to be broadly transferrable to 'activate' biological and synthetic soft materials, opening up an orthogonal route to pushing amorphous materials out of equilibrium. Applications of this class of topologically-active composites may include responsive suturing and wound-healing, self-healing infrastructure and programmable filtration and sequestration.

**Materials and Methods**

**DNA:** We prepare solutions of double-stranded DNA, 11 kilobasepairs (kbp) in length, *via* replication of pPIC11 plasmid constructs in *Escherichia coli* followed by extraction, purification and concentrating as described previously (*31*, *40*, *76*). Briefly, to replicate DNA, *E. coli* cultures containing the plasmid clone are grown from frozen glycerol stocks. To extract the DNA, cells are lysed via treatment with an alkaline solution. The extracted DNA is then renatured via treatment with an acidic detergent, precipitated in isopropanol, washed with 70% ethanol, and resuspended in nanopure deionized water (DI). To purify the DNA, the solution is treated with Rnase A (to remove contaminating RNA) followed by phenol-chloroform extraction and dialysis (to remove proteins). The purified DNA solution is further concentrated via rotary vacuum concentration and stored at 4°C.

We use gel electrophoresis and band intensity analysis to determine a concentration of 11.3 mg/mL of our stock circular DNA solution. Gel image analysis is performed using Life Technologies E-Gel Imager and Gel Quant Express software. The radius of gyration for the circular DNA is $R_{G,R} \simeq 196$ nm (*39*, *40*), from which we compute the polymer overlap concentration $c^*_{DNA} \simeq (3/4\pi)(M/N_A)R_{G,R}^{-3} \simeq 480$ µg/mL for the DNA solution (*11*, *35*, *39*). For all experiments, we dilute the stock DNA solution to $11c^* \simeq 5.28$ mg/ml in DI. The DNA length and buffer conditions are such that the DNA can be considered a flexible (rather than semiflexible) polymer (*40*, *44*, *77*, *78*).

**Dextran:** An aqueous solution of molecular biology grade dextran (Fisher BioReagents BP1580100, Lot #196289), with molecular weight 500 kDa, $R_G \simeq 19$ nm and $c^* \simeq 28.9$ mg/ml, is prepared by dissolving dextran in DI at a concentration of $11c^* \simeq 318$ mg/ml (*25*, *79*). Dextran is a randomly branched neutral polymer that assumes a random coil conformation in solution (*80–82*). Because the dextran radius of gyration is ~10x smaller than that of the DNA, we do not explicitly consider the effects of branching in our discussions and instead treat dextran as a neutral crowder with size ~19 nm ($R_G$). The solution is homogenized by mechanical agitation on a laboratory shaker at room temperature for >24 hrs.

**Restriction Endonuclease:** We use ApoI (New England BioLabs) as the restriction endonuclease that pushes the DNA solutions and composites out of equilibrium via topological conversion of DNA. We use the high fidelity (HF) version of the enzyme to ensure that no star activity occurs over the several hour digestion. ApoI cuts pPIC11 at 13 unique recognition sites, converting the circular DNA to linear topology (the first cut) and then cutting the resulting linear strand into 13 fragments of different lengths ranging from 64 bp to 3160 bp, with an average fragment length of $\langle l_f \rangle \simeq 854$ bp. This digestion can be seen using time-resolved gel electrophoresis as shown in Fig 1C and Fig 5, and described below (*21*).

**Sample preparation:** We prepare aqueous composite solutions comprising varying volume fractions of $11c^*$ solutions of DNA ($\phi_{DNA}$) and dextran ($\phi_{dx} = 1 - \phi_{DNA}$) (Fig 1A). Each sample has a total volume of 200 μL that includes 20 μL of 10× CutSmart Buffer (0.5 M Potassium Acetate, 0.2 M Tris-acetate, 0.1 M Magnesium Acetate; New England BioLabs) and DNA volume fractions of $\phi_{DNA} = 0, 0.25, 0.5, 0.75$ or 1. The buffer conditions and temperature (20°C) provide good solvent conditions for the DNA (*21, 40, 42–44*). Prior to each measurement, we mix and equilibrate the sample on a rotator at 4°C for 12-24 hours. Immediately before loading the sample on the rheometer, we mix in ApoI at an enzyme:DNA stoichiometry of 0.05 U/μg. We chose this relatively low stoichiometry to ensure that over the course of each rheological measurement ($t_m \simeq 13$ s), the systems can be treated as in quasi-steady state with respect to topological conversion. Specifically, our gel electrophoresis analysis (Fig 1C) shows that DNA digestion takes $t_d \gtrsim 270$ mins to complete in the $11c^*$ solutions and composites, giving a cutting rate of $t_c^{-1} \simeq 13$ cuts ÷ 270 min $\simeq 0.05$ min$^{-1}$ compared to a data acquisition rate of $t_m^{-1} \simeq (60 \text{ s/min}) \div 13 \text{ s} \simeq 5$ min$^{-1}$, resulting in a separation of timescales of $\chi \simeq (t_m^{-1}/t_c^{-1}) \sim 10^2$.

**Rheometry**: To perform bulk linear rheology measurements, we use a Discovery Hybrid Rheometer 3 (DHR3, TA Instruments) with 40 mm stainless steel parallel plate upper geometry and Peltier temperature-controlled bottom geometry fixed at 22°C. We load 180 μL of the sample onto the bottom plate and then lower the upper geometry until the gap is completely filled (~90 – 120 μm). To prevent evaporation during the experimental cycle we apply mineral oil around the geometry and the sample.

To measure the linear viscoelastic moduli, $G'(\omega)$ and $G'(\omega)$, over the course of enzymatic activity, we apply continuous oscillatory shear to the sample at a frequency of $\omega = 1$ rad/s and strain of $\gamma = 5\%$, acquiring data every $t_m \simeq 13$ s over the course of 6 hours (Fig 1D). We chose $\gamma = 5\%$ by performing amplitude sweeps and identifying the maximum strain that was still well within the linear regime. We chose $\omega = 1$ rad/s, via frequency sweeps, to be within the entanglement regime for the majority of the viscoelastic composites (Figs 2,3). Immediately following the 6-hr time sweep, we perform two consecutive frequency sweeps (Figure 1E) at $\gamma = 5\%$ for $\omega = 0.009 - 20$ rad/s to fully characterize the linear viscoelastic properties of the digested composites. We also perform identical time and frequency sweeps on samples without ApoI to characterize the viscoelastic properties of the composites in the absence of enzymatic activity. All data shown is above the measurable torque minimum of 0.5 nN·m for the DHR3 rheometer. We acquire and analyze data using TA Instruments TRIOS software and Origin Pro.

**Time-resolved Gel Electrophoresis:** To characterize the rate at which ApoI cleaves the 11 kbp circular DNA into linear fragments in the different DNA-dextran composite formulations, we use direct current agarose gel electrophoresis to separate the different topologies and lengths of DNA. Specifically, we prepare 40 μL samples of $11c^*$ DNA-dextran composite solutions that include 4 μL of 10× CutSmart Buffer, 0.05 U/μg of ApoI and DNA fractions of $\phi_{DNA} = 0.25, 0.5, 0.75$ or 1. We incubate each sample at RT for 6 hours, during which we remove a 1 μL aliquot from the reaction every 10 minutes and quench it with TE buffer and gel loading dye. We load 50 ng of DNA from each 'kinetic aliquot' onto a 1% agarose gel prepared with TAE buffer. We run each gel at 5 V/cm for 2.5 hours, allowing for separation of the DNA into distinct bands corresponding to ring and linear DNA of varying lengths. We use the standard $\lambda$-HindIII molecular marker ($\lambda$H) to calibrate the gel to determine the DNA topology and length corresponding to each distinct band (Figs 1C, 5).

**Single-molecule tracking:** To measure the dynamics of DNA molecules within the active composites we add a trace amount of fluorescent-labeled DNA to the composites. We label the DNA with covalent dye Mirus-488 at a dye:basepair ratio 1:5 using the Mirus Label IT Nucleic Acid Labeling Kit. We prepare sample chambers that accommodate ~10 μL by fusing together a glass coverslip and microscope with heated parafilm spacer. We passivate the surfaces with 10mg/mL BSA to prevent non-specific adsorption of the DNA. We use an Olympus IX73 inverted fluorescence microscope with a 60× 1.2 NA oil immersion objective (Olympus), 490/535 nm excitation/emission filter cube, and Hamamatsu Orca Flash CMOS camera to image the DNA. We collect a 2000-frame time-series of 1920 x 1440-pixel images at 33 fps every 10 minutes over the course of 6 hours (totaling 37 time-points) with $t = 0$ denoting the time that we add ApoI to the system. We use custom particle tracking scripts (Python) to track the center-of-mass of individual DNA molecules and measure their $x$ and $y$ displacements ($\Delta x, \Delta y$) for given lagtime, $\Delta t$. We compute van Hove distributions, i.e. probability distributions of particle displacements for a given $\Delta t$, by summing up the counts per displacement, dividing by the total number of counts, and binning the displacements into 150 equally-spaced intervals. To qualitatively depict the extent to which the distributions deviate from Gaussianity and exhibit heterogeneities, we fit each distribution to a Gaussian $P(\Delta x) = Ae^{-B(\Delta x)^2}$ where $A$ and $B$ are fitting parameters. To quantify the extent to which the displacements deviate from Gaussianity we compute the non-Gaussian parameter $NG = \frac{1}{3}\frac{\langle \delta^4(\Delta t)\rangle}{\langle \delta^2(\Delta t)\rangle^2} - 1$ and ergodicity breaking parameter $EB = \frac{\langle(\overline{\delta^2(\Delta t)})^2\rangle - \langle\overline{\delta^2(\Delta t)}\rangle^2}{\langle\overline{\delta^2(\Delta t)}\rangle^2}$ where $\delta^2(\Delta t)$ is the squared displacement of individual DNA molecules and $\langle \cdot \rangle$ and $\overline{\phantom{\cdot}}$ denote ensemble and time averaging, respectively (*74*, *75*) over the course of the 6-hr digestion. Each data point plotted in Fig 6H,I is an average over all lag times in a given time-series. The van Hove distributions are plotted for $\Delta t = 1.21$ s, and are representative of the majority of distributions computed across sampled lag times.

**Acknowledgments**

We acknowledge P. Khanal for assistance in sample preparation and rheology protocols, and H.H. Winter for insightful discussions.

**Funding:**

Air Force Office of Scientific Research grant AFOSR- FA9550-17-1-0249 (RMRA)
National Institutes of Health grant NIH NIGMS R15GM123420 (RMRA, RJM, JM)
National Science Foundation grant NSF-DMR-2203791 (RJM, RMRA)
National Science Foundation grant NSF-CHE- 2050846 (JM, RMRA)

**Author contributions:**

Conceptualization: RMRA, RJMG
Methodology: RMRA, RJMG, PN
Investigation: JM, RMG
Visualization: JM, RMRA
Supervision: RMRA, RJMG
Writing—original draft: JM, RMRA
Writing—review & editing: JM, RMRA, RJMG

**Competing interests:** Authors declare that they have no competing interests.

**Data and materials availability:** All data needed to evaluate the conclusions in the paper are present in the paper, Supplementary Materials, or the Open Science Framework url: osf.io/2jvea.

# Science Advances

Supplementary Materials for

**Cooperative rheological state-switching of enzymatically-driven composites of circular DNA and dextran**

Juexin Marfai *et al.*

Corresponding author. Email: randerson@sandiego.edu

**This PDF file includes:**

**Figure S1.** Effect of ApoI digestion on the frequency-dependent loss tangents $\tan\delta(\omega)$, for DNA-dextran composites with varying $\phi_{DNA}$.

**Figure S2.** Time-dependent viscous modulus $G''(t_a)$, loss tangent $\tan\delta(t_a)$, and complex viscosity $\eta^*(t_a)$ for DNA-dextran composites during ApoI digestion.

**Figure S3.** Effect of ApoI digestion on the frequency-dependent complex viscosity $\eta^*(\omega)$ and zero-shear viscosity $\eta_0$ of DNA-dextran composites with varying $\phi_{DNA}$.

**Table S1.** Rheological parameters determined by evaluating (A) $G'(t_a)$ and (B) $\tan\delta(t_a)$ data and corresponding fits.

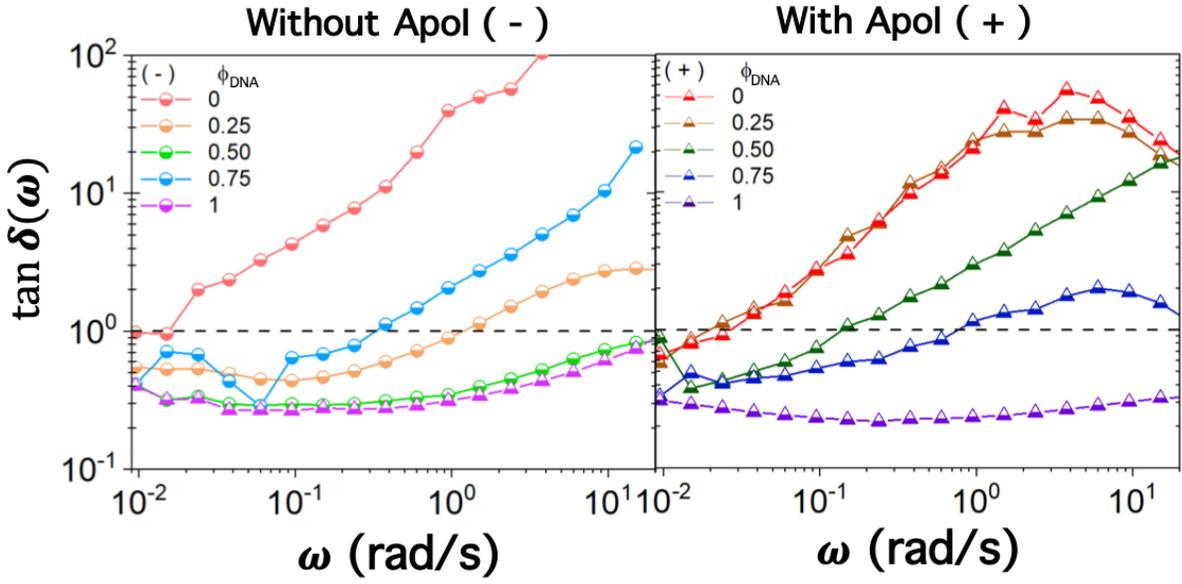

**Fig. S1. Effect of ApoI digestion on the frequency-dependent loss tangents for DNA-dextran composites with varying $\phi_{DNA}$.** $\tan\delta(\omega)$ versus $\omega$, computed as $G''(\omega)/G'(\omega)$ from the data shown in Figs 2a and 3(a-c), measured after 6 hours of either ApoI digestion (+, right) or under identical conditions but with no ApoI (-, left). The dashed horizontal line indicates $\tan\delta = 1$ that crosses each data set at its corresponding crossover frequency $\omega_c$. Colors and symbols are indicated in the legend and match those of Figs 2 and 3.

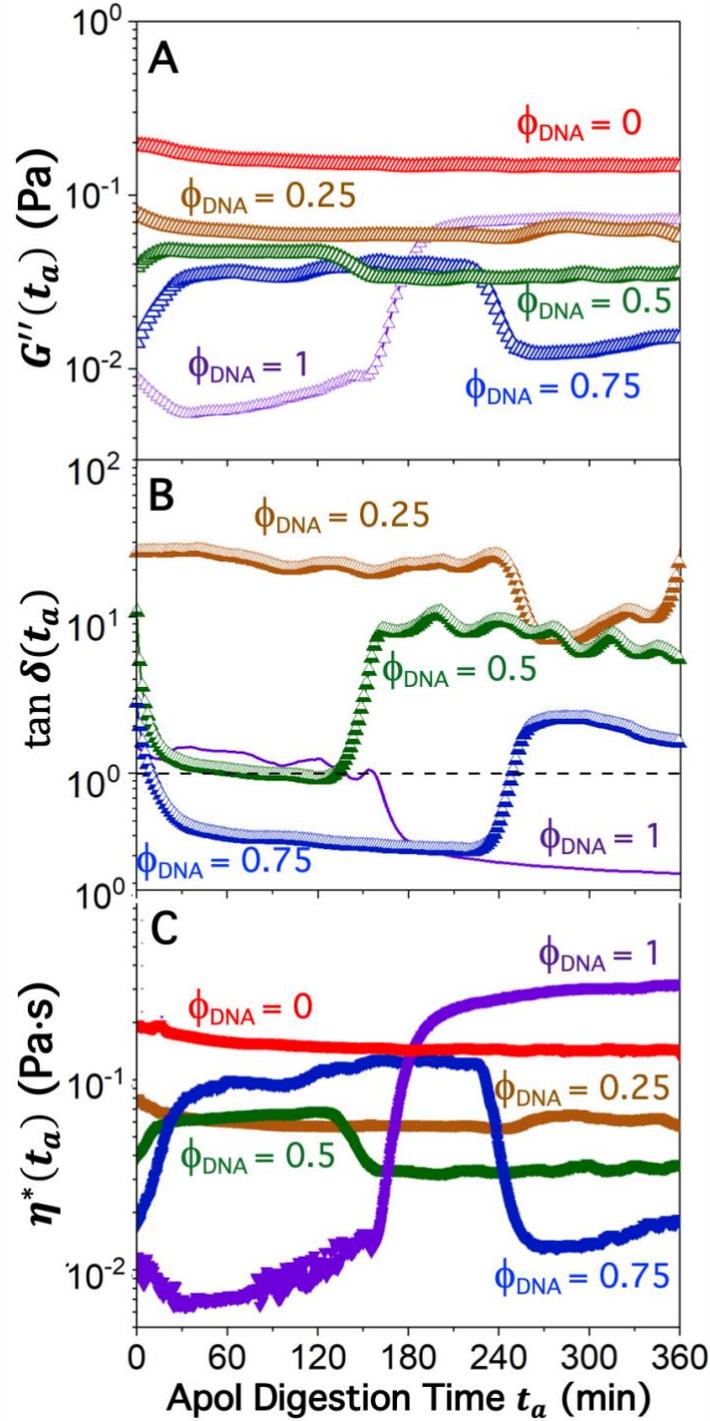

**Fig. S2. Time-dependent viscous modulus $G''(t_a)$, loss tangent $\tan\delta(t_a)$, and complex viscosity $\eta^*(t_a)$ of DNA-dextran composites during ApoI digestion.** (A) $G''(t_a)$, (B) $\tan\delta(t_a)$, and (C) $\eta^*(t_a)$ curves correspond to the same data as shown in Figs 2 and 4 for $\phi_{DNA} = 0$ (red), $0.25$ (orange), $0.5$ (green), $0.75$ (blue), and $1$ (purple).

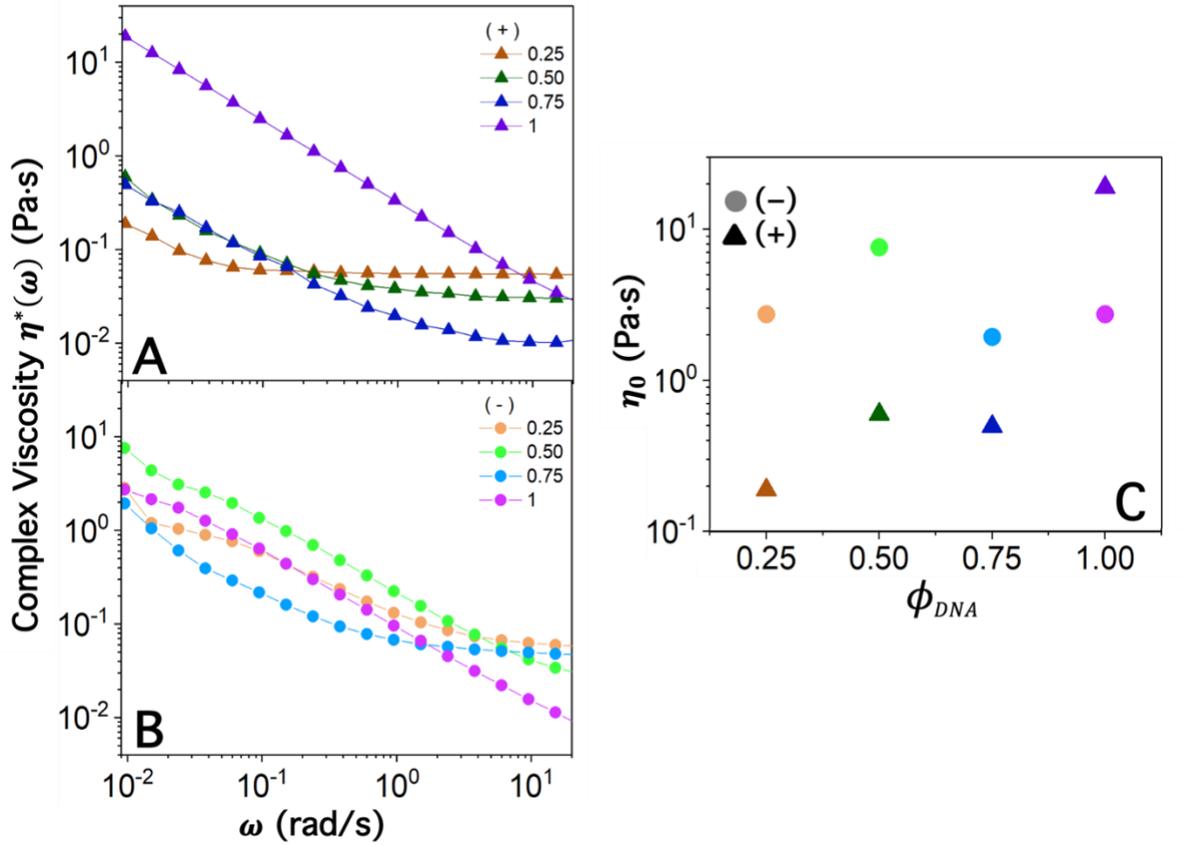

**Fig. S3. Effect of ApoI digestion on the frequency-dependent complex viscosity $\eta^*(\omega)$ and zero-shear viscosity $\eta_0$ of DNA-dextran composites with varying $\phi_{DNA}$. (A,B)** $\eta^*(\omega)$ versus $\omega$, computed as $\eta^*(\omega) = \left[\left(G'(\omega)\right)^2 + \left(G''(\omega)\right)^2\right]^{1/2} \omega^{-1}$ from the data shown in Figs 2a and 3(a-c), measured after 6 hours of either (A) ApoI digestion (+, triangles) or (B) under identical conditions but with no ApoI (-, circles). **(C)** Zero-shear viscosity $\eta_0$ estimated as the low-frequency limit of the $\eta^*(\omega)$ data shown in A,B. Note that $\eta_0$ values are lower-bounds as the complex viscosities do not reach $\omega$-independent plateaus. Values are used to estimate the structural relaxation times $t_0$ shown in Fig 3.

### A

| $\phi_{DNA}$ | Source | $G'_i$ (mPa) | $G'_h$ (mPa) | $G'_l$ (mPa) | $\tau_1$ (min) | $\tau_2$ (min) | $n_1$ | $n_2$ |
|---|---|---|---|---|---|---|---|---|
| 0.25 | experiment | 2.69 | 7.4 | 2.65 | 255 | 355 | - | - |
|  | fit | 2.41 | 9.13 | 1.19 | 259 | 355 | 61.3 | 101.3 |
| 0.5 | experiment | 3.91 | 46.9 | 4.46 | 7.5 | 148 | - | - |
|  | fit | *5.39* | 46.6 | *3.55* | 11.4 | 142 | 1.7 | 28 |
| 0.75 | experiment | 7.42 | 106 | 6.85 | 13.5 | 243 | - | - |
|  | fit | *9.74* | 95.2 | *5.37* | *23.1* | 236 | 2.47 | 47.7 |
| 1 | experiment | 5.27 | 283 | - | 168 | - | - | - |
|  | fit | *5.95* | 228 | - | 182 | - | 22.4 | - |

### B

| $\phi_{DNA}$ | Source | $\tan\delta_i$ | $\tan\delta_h$ | $\tan\delta_l$ | $\tau_1$ (min) | $\tau_2$ (min) | $n_1$ | $n_2$ |
|---|---|---|---|---|---|---|---|---|
| 0.25 | experiment | 22.6 | 8.86 | 21.7 | 236 | 353 | - | - |
|  | fit | 23.4 | *6.96* | *23.9* | 254 | 356 | 65.4 | 100 |
| 0.5 | experiment | 9.79 | 0.99 | 7.8 | 3.3 | 149 | - | - |
|  | fit | 12 | 0.82 | 8.2 | 1.32 | 153 | 1.09 | 45 |
| 0.75 | experiment | 2.06 | 0.35 | 1.99 | 3.07 | 247 | - | - |
|  | fit | 2.79 | 0.31 | 2.22 | 3.99 | 251 | 1.19 | 53.4 |
| 1 | experiment | 1.29 | 0.25 | - | 166 | - | - | - |
|  | fit | 1.07 | *0.32* | - | 164 | - | 35.4 | - |

**Table S1. Rheological parameters determined by evaluating (A) $G'(t_a)$ and (B) $\tan\delta(t_a)$ data and corresponding fits**. For each composite composition the values determined directly from the experimental data and fit to effective Hill function (A) $G'(t_a) = (G'_{h/l} - G'_{i/h})/(1 + \left(\frac{\tau_{1/2}}{t_a}\right)^{n_{1/2}}) + G'_{i/h}$ and (B) the equivalent for $\tan\delta(t_a)$ for the first/second state transition regions (see Fig 5 for fits to $G'(t_a)$). Cells that are empty are because (1) $\phi_{DNA} = 1$ does not reach a second transition, and (2) cooperativity values $n_{1/2}$ are fit parameters for which there is no straightforward method for determining without fitting. The standard error associated with each average value listed is <1% of the average value except for the italicized values that have percent errors of <5% (and >1%). Errors from fit values are the error associated with the fit, and errors for experimental values are the standard error of values across each time-independent plateau regime.